\documentclass[11pt,amssymb,amsfonts,amscd]{article}
\usepackage{amscd}
\usepackage{amssymb}
\usepackage{amsmath}
\usepackage{amsfonts}
\newcommand{\nn}{\nonumber}

\def\be{\begin{equation}}
\def\ee{\end{equation}}

\newcommand{\CC}{{\mathbb C}}
\newcommand{\RR}{{\mathbb R}}
\newcommand{\ZZ}{{\mathbb Z}}

\newcommand{\QQ}{{\mathbb Q}}

\newcommand{\ra}{\rightarrow}

\newcommand{\Ext}{\mathbf{Ext}}

\newcommand{\op}{\oplus}
\newcommand{\ot}{\otimes}

\newcommand{\bz}{{\bar{z}}}

\newcommand{\bpartial}{{\bar{\partial}}}

\newcommand{\cO}{{\mathcal O}}

\newcommand{\cT}{{\mathcal T}}

\newcommand{\cF}{{\mathcal F}}

\newcommand{\cL}{{\mathcal L}}

\newcommand{\bpsi}{{\bar\psi}}

\newcommand{\bQ}{{\bar Q}}
\newcommand{\bJ}{{\bar J}}
\newcommand{\cW}{{\mathcal W}}

\newcommand{\lto}{\longrightarrow}
\newcommand{\ts}{\times}
\newcommand{\fu}{\phi}
\newcommand{\Pic}{{\rm P}{\rm i}{\rm c}}

\title{Remarks On A-branes, Mirror Symmetry, And The Fukaya Category}

\author{Anton Kapustin\thanks{%
California Institute of Technology,
Pasadena, CA 91125,
E-mail: kapustin@theory.caltech.edu}
\and Dmitri Orlov%
\thanks{%
Algebra Section, Steklov Mathematical Institute,
Russian Academy of Sciences,
 8 Gubkin str., GSP-1, Moscow 117966, Russia,
E-mail: orlov@mi.ras.ru}}
\date{}
\begin{document}

\begin{titlepage}

\maketitle

\begin{abstract}

We discuss D-branes of the topological A-model (A-branes), which
are believed to be closely related to the Fukaya category.
We give string theory arguments which show that A-branes are
not necessarily Lagrangian submanifolds in the Calabi-Yau: more
general coisotropic branes are also allowed, if the line bundle on
the brane is not flat. We show that a coisotropic A-brane has a natural
structure of a foliated manifold with a transverse holomorphic structure. We argue that the Fukaya category must be
enlarged with such objects for the Homological Mirror Symmetry
conjecture to be true.

\noindent

\end{abstract}
\vspace{-4.8in}

\parbox{\linewidth}
{\small\hfill \shortstack{\large CALT-68-2348}}

\end{titlepage}

\section{Introduction}\label{intro}

Let $X$ be a weak Calabi-Yau manifold, i.e. a complex manifold with $c_1(X)=0$ 
which admits a K\"ahler metric. Given a Ricci-flat K\"ahler metric $G$ on 
$X,$ and a B-field (a class in $H^2(X,\RR))$, one can canonically construct an $N=2$ supersymmetric sigma-model with ``target'' $X.$
On physical grounds, the quantized
version of this model has $N=2$ superconformal symmetry and describes propagation of closed strings on $X.$ In this note we set $B=0$ for simplicity. According to Calabi's conjecture proved by Yau, we can parametrize $G$ by the cohomology class of its K\"ahler form
$\omega.$ A weak Calabi-Yau manifold equipped with a K\"ahler form $\omega$ will be called a physicist's Calabi-Yau.

It sometimes happens that two different physicist's Calabi-Yau
manifolds $(X,\omega)$ and $(X',\omega')$ give rise to a pair of $N=2$
superconformal field theories (SCFTs) related by a {\it mirror morphism}~\cite{GrPl,CdelaOGP}. 
A mirror morphism of $N=2$ SCFTs is an isomorphism of the underlying $N=1$
SCFTs which acts on the $N=2$ super-Virasoro algebra as a mirror involution~\cite{Dixon,LVW}.
In this case one says that
$(X,\omega)$ and $(X',\omega')$ are mirror to each other.
(For a concise explanation of the notions involved and further references
see~\cite{V}. An algebraically--minded reader may
find it useful to consult Ref.~\cite{KO} for a careful definition of
$N=2$ SCFTs and their morphisms.)

A long-standing problem is to understand the mirror relation from a mathematical viewpoint, i.e. without a recourse to
the ill-defined procedure of quantizing a sigma-model.
A fascinating conjecture has been put forward by
M.~Kontsevich~\cite{HMSC}. He observed that
to any physicist's Calabi-Yau $(X,\omega)$ one can associate two
triangulated categories: the well-known bounded derived category of coherent sheaves $D^b(X)$ and the still mysterious Fukaya category
$D\cF(X).$ Objects of the category $D^b(X)$ are bounded complexes
of coherent sheaves. Objects of the Fukaya category
are (roughly speaking) vector bundles on Lagrangian submanifolds
of $X$ equipped with unitary flat connections.
The Homological Mirror Symmetry Conjecture (HMSC) asserts~\cite{HMSC}
that if two algebraic physicist's Calabi-Yau manifolds
$(X, \omega)$ and $(X', \omega')$ are mirror to each other,
then $D^b(X)$ is equivalent to $D\cF(X'),$ and $D\cF(X)$ is equivalent
to $D^b(X').$
So far this conjecture has been proved only for elliptic curves~\cite{PZ}.

{}From a physical viewpoint, complexes of coherent sheaves are
D-branes of the topological B-model (B-branes). We remind that the B-model of a physicist's Calabi-Yau $(X,\omega)$ is a topological ``twist'' of the corresponding $N=2$ SCFT~\cite{Wm}. 
The twisted theory is
a two-dimensional topological field theory whose correlators
do not depend on $\omega.$ Morphisms between
the objects of $D^b(X)$ are identified with the states of the
topological string stretched between pairs of B-branes, and the compositions of morphisms are computed by the correlators of the B-model. This correspondence has been intensively discussed
in the physics literature (see for example \cite{Du,Laz,AL,Dia}
and references therein), and will be taken as a starting point here.

An $N=2$ SCFT has another twist, called the A-twist~\cite{Wm}. 
The corresponding topological field theory (the A-model) is insensitive
to the complex structure of $X,$ but depends non-trivially on the
symplectic form $\omega.$ D-branes of the A-model are called A-branes. Mirror morphisms exchange A- and B-twists and A- and B-branes. Thus
from a physical viewpoint the mirror of $D^b(X)$ is the
category of A-branes on $X'.$

It can be shown that any object of the Fukaya category gives rise
to an A-brane. Moreover, the recipe for computing morphisms between
such A-branes can be derived heuristically in the path integral formalism, and it reproduces the definition of morphisms in the
Fukaya category~\cite{Witten}.
Therefore the majority of researchers in the
field assumed that the mirror relation between the categories of
A- and B-branes is essentially a restatement of the HMSC in physical
terms.\footnote{In fact, the calculation of morphisms between
Lagrangian A-branes in Ref.~\cite{Witten} preceded the formulation of 
the HMSC and served as an important motivation for it.}

In this note
we will argue that this is not the case, because A-branes are
not necessarily Lagrangian submanifolds in $X.$
This was mentioned already in one of the first papers on the 
subject~\cite{OOY}, but the general conditions for a D-brane
to be an A-brane have not been determined there.
In Section 3 we will show
that a coisotropic submanifold of $X$ with a unitary line bundle on
it is an A-brane if the curvature of the connection satisfies
a certain algebraic condition. We remind that a submanifold $Y$ of
a symplectic manifold $(X,\omega)$ is called coisotropic if the
skew-complement of $TY\subset TX\vert_Y$ with respect to $\omega$ 
is contained in $TY.$ In the physical language, a coisotropic 
submanifold is a submanifold locally defined by first-class constraints.
One can easily see that the dimension of a coisotropic submanifold is at least half the dimension of $X,$
and that a middle-dimensional coisotropic submanifold is
the same thing as a Lagrangian submanifold. Thus we show that
the category of A-branes contains, besides Lagrangian A-branes,
A-branes of larger dimension.

In Section 4 we explore the
geometric interpretation of the algebraic condition on the
curvature of the line bundle. We will see that
an A-brane is naturally a {\it foliated manifold with a transverse
holomorphic structure}. The notion of transverse holomorphic structure 
is a generalization of the notion of complex structure to foliated
manifolds. If the space of leaves of a foliated
manifold $Y$ is a smooth manifold, a transverse holomorphic structure 
on $Y$ is simply a complex structure on the space of leaves.
The general definition is given in Section~4. In addition to being
transversely holomorphic, a coisotropic A-brane also carries a
transverse holomorphic symplectic form.

In the case of a Lagrangian A-brane, the foliation has codimension zero, there are no transverse directions, and the transverse holomorphic
structure is not visible. In general, the foliation is determined by the restriction of $\omega$ to $Y$, while the transverse holomorphic
structure comes from the curvature of the line bundle on the brane.

Interestingly, to prove that an A-brane has a natural transverse
holomorphic structure, one needs to use some facts from
bihamiltonian geometry. The subject matter of bihamiltonian geometry
is manifolds equipped with two compatible (in a sense explained below) Poisson structures. In our case, the underlying manifold is foliated, and one is dealing with {\it transverse} Poisson structures.
(If the space of leaves is a manifold, specifying a transverse Poisson
structure is the same as specifying an ordinary Poisson structure on the space of leaves.)
One of the transverse Poisson structures arises from the symplectic
form $\omega$ in the ambient space $X,$ and the other one from the
curvature of the line bundle on $Y.$ 

Our understanding of the category of A-branes is far from complete. Nevertheless, it is clear that generally it includes objects other than Lagrangian submanifolds with flat vector bundles. 
(There are certain special, but important, cases where there seem to be
no non-Lagrangian A-branes, like the case of an elliptic curve, or a simply-connected Calabi-Yau 3-fold.) Therefore 
the Fukaya category must be enlarged with coisotropic A-branes for the HMSC to be true. (This is somewhat reminiscent of the remark
made in~\cite{HMSC} that Lagrangian foliations may need to be included
in the Fukaya category.) This is discussed in more detail in Section~5.

Since our arguments are ultimately based on non-rigorous physical
reasoning, a skeptic might not be convinced that the HMSC needs
serious modification. To dispel such doubts, we discuss in
Section~2 mirror symmetry for tori and show that under mild
assumptions the usual Fukaya category cannot capture the subtle behavior
of $D^b(X)$ under the variation of complex structure. Inclusion
of coisotropic A-branes seems to resolve the problem.

\section{Why Lagrangian submanifolds are not enough}

In this section we give some examples
which show that the Fukaya category must be enlarged with non-Lagrangian objects for the HMSC to be true. We will
exhibit a mirror pair of tori such that mirror symmetry takes a holomorphic line bundle (a B-brane) on the first torus to a complex line bundle on the second torus. This means that the latter line bundle is an A-brane.

It is well known that the derived category of coherent sheaves
behaves in a very non-trivial manner under a variation of complex structure, and at special loci in the moduli space of complex
structures it can become ``larger.''
This is easy to see on the level of the Grothendieck group of
$D^b(X),$ which we denote by $K_0(D^b(X)).$ There is a map
$$
ch: K_0(D^b(X))\ot \QQ \lto H^*(X, \QQ)
$$
called the Chern character. The image of this map is contained in
the intersection of $H^*(X, \QQ)$ with $\oplus_p H^{p,p}(X)$ in
the complex cohomology group $H^*(X, \CC)$ and, by the Hodge
Conjecture, should coincide with this intersection.

Let us denote by $NS(X)$ the Neron-Severi group of $X$ which,  by definition, is the image of a natural map from the Picard group $\Pic(X)$ to $H^2(X,\ZZ).$ Then we have
$NS(X)\ot \QQ=Im(ch)\bigcap H^2(X, \QQ),$ and therefore
$Im(ch)$ contains a subring generated by the Neron-Severi
group.

One can see from examples that the image of the map $ch$ can
change under a variation of complex structure; in particular, the
dimension of $Im(ch)$ can jump if, for example, the dimension of
the Neron-Severi group jumps.

The ``jumping'' phenomenon can be easily observed in the case of
abelian varieties. Let $E_{\tau}$ be an elliptic curve with
a Teichm\"uller parameter $\tau.$
It  has a structure of an algebraic group.
Let $e$ be the identity point of this group. 
It can be checked that any
endomorphism of $E_{\tau}$ that sends the point $e$ to itself is an
endomorphism of the algebraic group. Such endomorphisms form a ring
which contains $\ZZ$ as a subring and for a ``generic'' elliptic curve
coincides with it. However the ring of $e-$\!\!preserving 
endomorphisms of
$E_{\tau}$ can be bigger than $\ZZ.$ In this case one says that the
elliptic curve $E_\tau$ possesses a complex multiplication. 
It can be shown that $E_{\tau}$ has a complex multiplication iff $\tau$ is a root of a quadratic polynomial with integral coefficients. For example, the elliptic curve with $\tau=i$ is an example of a curve with 
a complex multiplication. 

Let $E_{\tau}$ be an elliptic curve with a complex multiplication.
Consider an $n$-dimensional abelian variety $A=E^n_{\tau}$ with $n\ge 2.$ 
In this case the derived category $D^b(A)$ is in a certain sense much
bigger than the derived category of a ``generic'' abelian variety.
For a ``generic'' abelian variety the Neron-Severi group is $\ZZ$ and, moreover, $NS(A)\ot \QQ$ generates the whole $Im(ch).$ 
Thus the dimension
of $Im(ch)$ is equal to $n+1.$ For an abelian variety $E_{\tau}^n,$
where $E_{\tau}$ is a ``generic'' elliptic curve, the dimension of
the Neron-Severi group is $n(n+1)/2.$ If the elliptic curve posesses 
a complex multiplication, then $\dim NS(A)=n^2$ and, moreover, we 
have an equality
$$
Im(ch)\ot \CC=\bigoplus_p H^{p,p}.
$$
Thus in this case $\dim_{\QQ}Im(ch)=\binom{2n}{n}.$

For example, for $n=2,$ if $\tau$ is generic,
the Neron-Severi group has dimension $3$ and is generated by
the divisors $\{pt\}\times E_{\tau},\, E_{\tau}\times \{pt\},\,
\Delta,$ where $\Delta$ is the diagonal of $E_\tau\ts E_\tau.$
In contrast, when $E_\tau$ posesses complex multiplication,
$NS(A)$ has dimension $4,$ which coincides with the dimension 
of $H^{1,1}(A).$ It is generated by the divisors
$\{pt\}\times E_{\tau},\, E_{\tau}\times \{pt\},\, \Delta,\, \Gamma,$
where  $\Gamma\subset E_\tau\times E_\tau$ is the graph of 
an additional endomorphism of $E_{\tau}.$

Now let us look at the Fukaya category of a mirror torus.
The mirror relation for abelian varieties is
well-understood~\cite{GLO,KO} (see also \cite{Ma}).
In particular, it is known that for any abelian variety $A$ one can
find a symplectic form $\omega$ such that for the pair
$(A, \omega)$ there exists a mirror-symmetric  abelian variety $B$
with a symplectic form $\omega_{B}$ (\cite{GLO}, Prop. 9.6.1).
Let $D\cF(B, \omega_{B})$ be the Fukaya category of the symplectic
manifold $(B, \omega_{B}).$
This category essentially depends only on the symplectic form $\omega_{B}$
and does not depend on the complex structure of the variety $B.$
This is mirror to the obvious fact that
the derived category of coherent sheaves does not depend
on the symplectic form.
By the HMSC the category $D\cF(B, \omega_{B})$
should be equivalent to the derived category
$D^b(A).$

Furthermore, the mirror correspondence induces an
isomorphism of the cohomology vector spaces
$$
\beta: H^*(A, \QQ)\stackrel{\sim}{\lto} H^*(B, \QQ).
$$
For abelian varieties the isomorphism $\beta$ is described in \cite{GLO}. It is natural to assume that $\beta$ is compatible with
the conjectured equivalence between the derived category $D^b(A)$ and the Fukaya category $D\cF(B, \omega_{B}).$
This means that there should exist a map $\fu$ from the Grothendieck  group $K_0(D\cF(B, \omega_{B}))\ot \QQ$ to the cohomology group $H^*(B, \QQ)$ which closes the commutative diagram
$$
\begin{CD}
K_0(D^b(A))\ot \QQ& @>\sim>>& K_0(D\cF(B, \omega_{B}))\ot \QQ\\
@V{ch}VV && @VV{\fu}V\\
H^*(A, \QQ)&@>{\beta}>>& H^*(B, \QQ)
\end{CD}
$$
Under the map $\fu$ a flat vector bundle on a Lagrangian submanifold
goes to the corresponding cycle in the middle-dimensional
cohomology group $H^n(B, \QQ)$ with a multiplicity equal to the rank of the bundle.

Now note that classes of Lagrangian submanifolds in the
middle-dimensional cohomology group belong to the kernel of a
surjective map
$$
H^n(B, \CC)\stackrel{\cdot [\omega_{B}]}{\lto} H^{n+2}(B, \CC).
$$
The dimension of the kernel is equal to $\binom{2n}{n}-\binom{2n}{n+2},$
which is less than the dimension of $Im(ch).$
Therefore when $A=E_\tau^n,$ where $E_\tau$ is an elliptic curve with
a complex multiplication, Lagrangian submanifolds in $B$ with flat 
vector bundles can not generate the mirror of $D^b(A),$ in contradiction with the HMSC.

To obtain some information on the mysterious mirror of $D^b(A),$
let us describe the mirror symmetry correspondence
for $A=E_{\tau}^n$ more explicitly. In this case mirror symmetry
is a T-duality. For simplicity we let $\tau=i,$ so that $E_\tau$ is
a ``square torus.''
Consider a decomposition of the lattice
$H^1(A, \ZZ)=\varGamma\op \varSigma$
with bases $\varGamma=\langle x_1,...,x_n \rangle$
and $\varSigma=\langle y_1,....,y_n\rangle$
such that the complex structure $I_A$ takes $x_i$ to $y_i$ and $y_i$ to $-x_i.$
Let $\langle l_1,..., l_n\rangle$ be the dual basis in
the dual lattice $\varGamma^*.$ A mirror manifold for the
abelian variety $A$ can be constructed by T-dualizing the directions
$x_1,\ldots,x_n.$ This means that the mirror manifold $B$ is a
torus $(\varGamma^*\op\varSigma)\ot \RR/(\varGamma^*\op\varSigma)$ equipped with a constant symplectic form
$$
\omega_{B}=\sum_{i=1}^{n} l_i \wedge y_i.
$$
(For simplicity we do not introduce
a symplectic form on $A$ and a complex structure on $B.)$

In this case the map $\beta$ is defined in the following way.
Let $T$ be a real $3n$-dimensional torus $\varPi\ot \RR/\varPi,$
where $\varPi=\varGamma\op\varSigma\op\varGamma^*.$
The torus $T$ has natural projections $p$ and $q$ to the tori $A$ and $B$:
$$
\begin{CD}
T&@>q>>&B\\
@VpVV&&\\
A&&
\end{CD}
$$
Let $P$ be a complex line bundle on $T$ defined by its first
Chern class:
$$
c_1(P)=\sum_{i=1}^{n} x_i\cdot l_i .
$$
The Chern character $ch(P)\in H^*(T, \QQ)$
is equal to $\exp(c_1(P)).$
According to~\cite{GLO}, the map $\beta$ from $H^*(A, \QQ)$ to
$H^*(B, \QQ)$ is given by the formula
$$
\beta(a):= q_*(ch(P)\cdot p^*(a)).
$$
(To define the map $q_*$ we chose fundamental classes
of $T$ and $B$ and used the Poincare duality between cohomology
and homology groups).
Using this formula, one can explicitly calculate
the subspace $\beta(Im(ch)).$

To demonstrate the existence of objects in the mirror of
$D^b(A)$ which are not Lagrangian submanifolds, we let $n=2$
for simplicity and consider a holomorphic line bundle $L$ on $A$
whose first Chern class is equal to
$$
c_1=x_1\cdot x_2 + y_1\cdot y_2.
$$
Such a holomorphic line bundle exists because $c_1\in H^{1,1}(A).$
The moduli space of such holomorphic line bundles
is a homogeneous space over $\Pic^0(A),$ the kernel of the natural
map from $\Pic(A)$ to $NS(A).$
More explicitly, $L$ is of the form $\cO(-D)\ot N,$ where $N\in
\Pic^0(A),$ $D=\Gamma-\{pt\}\times E - E\times \{pt\},$ and $\Gamma$
is the graph of the automorphism of $E$ given by multiplication
by $i.$
A direct calculation shows that
$$
\beta(ch(L))=\left(1;\; y_1\cdot y_2 -l_1 \cdot l_2; \;
-y_1\cdot y_2\cdot l_1\cdot l_2 \right)
\in H^{even}(B, \QQ).
$$
We see that $\beta(ch(L))$ coincides with the Chern character
$ch(M)$ of a complex line bundle $M$ on $B$ with the first Chern
class equal to
$$
c_1(M)=y_1\cdot y_2 - l_1\cdot l_2.
$$
Therefore it is natural to expect that the complex line bundle $M$
(with an unitary connection)
is an object of the mirror of $D^b(A),$ and that the
invertible coherent sheaf $L$ goes to the line bundle $M$ under
the mirror symmetry correspondence described above.
In physical terms, this
shows that the mirror of a D4-brane of type B with a flux
wrapped on a 4-torus can be a D4-brane of type A with a flux
wrapped on the mirror torus.

One can check that in this case the subspace $\beta(Im(ch))$
consists of the elements $(r; c; s)\in H^{even}(B, \QQ)$ such that
\begin{equation}\label{condchern}
c\cdot \omega_{B}=0,
\qquad
s=\frac{1}{2}r{\omega_{B}}^2.
\end{equation}

Similarly, for any $n>2$ we can find elements of $\beta(Im(ch))$
which do not belong to the middle cohomology group of $B$ and
therefore correspond to non-Lagrangian objects of the mirror
of $D^b(A).$

One may ask how general this phenomenon is.
It does not occur for odd-dimensional 
Calabi-Yau manifolds which are complete intersections in 
projective spaces.
But it seems that for even-dimensional Calabi-Yaus
(for example, for K3 surfaces) or for more general odd-dimensional
Calabi-Yaus the situation is similar
to that for abelian varieties,
i.e. non-Lagrangian A-branes appear at special points
in the moduli space of symplectic structures.

\section{World-sheet approach to A-branes}

This section assumes some familiarity with supersymmetric
sigma-models (on the
classical level) and superconformal symmetries. Let $X$ be a K\"ahler
manifold with metric $G$ and K\"ahler form $\omega.$
The complex structure on $X$ is given by $I=G^{-1}\omega.$ The supersymmetric
sigma-model with target $X$ classically has $(2,2)$
superconformal symmetry. Quantum anomaly destroys this symmetry
unless $c_1(X)=0.$

Let $j:Y\ra X$ be a submanifold in $X,$ and $E$ be a line bundle on $Y$ with a
unitary connection. Our goal is to derive the necessary and sufficient conditions
for a pair $(Y,E)$ to be a D-brane of type A. We will find that these
conditions depend on $\omega,$ but are not sensitive to the complex structure
on $X,$ as expected on general grounds.

Let $\cW$ be an open string world-sheet, i.e. a Riemann surface with a boundary.
The fields of the sigma-model consist of a smooth map $\Phi:\cW\ra X,$ and
sections $\psi,\bpsi$ of $\Phi^*(TX)\ot \Pi S^\pm.$ Here $S^\pm$ are
semi-spinor line bundles
on $\cW,$ and $\Pi$ is the parity-reversal functor. In the physical
language, $\Phi$ is a bosonic field, while $\psi$ and $\bpsi$
are fermionic fields. The precise form of the action
is unimportant for our purposes; what is important is that the action has $(2,2)$ superconformal symmetry. In particular, the supercurrents $Q^\pm,\bQ^\pm$ are given by
\begin{eqnarray}\nn
Q^\pm &=& \frac{i}{4\sqrt 2} G\left(\psi,\partial \Phi\right)\pm
\frac{1}{4\sqrt 2}\omega\left(\psi,\partial \Phi\right),\\ \nn
Q^\pm &=& \frac{i}{4\sqrt 2} G\left(\bpsi,\bpartial \Phi\right)\pm
\frac{1}{4\sqrt 2}\omega\left(\bpsi,\bpartial \Phi\right),
\end{eqnarray}
and the $U(1)$ R-currents are given by
\begin{eqnarray}\nn
J &=& -\frac{i}{2}\omega\left(\psi,\psi\right),\\ \nn
\bJ &=& -\frac{i}{2}\omega\left(\bpsi,\bpsi\right).
\end{eqnarray}
Supercurrents and R-currents are sections of powers of the semi-spinor
bundles.

Consider open strings ending on $Y,$ i.e. maps $\Phi$ such that some or all of the
components of $\partial\cW$ are mapped to $Y.$ For example, we may consider
the situation where $\cW$ is an upper half-plane, and $\partial\cW$ is
the real axis. Then the map $\Phi$ and the sections $\psi,\bpsi$ must satisfy on the boundary $z=\bz$ the following conditions:
\begin{eqnarray}
\partial \Phi&=&R\left(\bpartial\Phi\right),\label{reflb}\\
\psi&=&R(\bpsi).\label{reflf}
\end{eqnarray}
Here $R$ is an endomorphism of the restriction of $TX$ to $Y.$ Furthermore,
$R$ can be expressed in terms of $G$ and the curvature of the line bundle $E.$
To write it down, we will use the metric $G$ to decompose $TX\vert_Y$ as
$NY\op TY.$
$R$ preserves this decomposition and has the form
\begin{equation}\label{R}
R=\left(-id_{NY}\right)\op (g-F)^{-1}(g+F).
\end{equation}
Here $g$ is the restriction of $G$ to $Y,$ and $F$ is the curvature 2-form of
the line bundle $E.$ (We use the physical convention in which $F$ is real.)

The physical meaning of this formula is very simple. Recall that the boundary
of the string world-sheet $\cW$ is the trajectory of
a string end-point, and that the string end-point is charged with respect to
the gauge field on the brane~\cite{Polchinsky}.
Thus for non-zero $F$ there is a Lorenz
force acting on the end-point. Eqs.~(\ref{reflb}) and (\ref{R}) say that the velocity of the end-point is tangent to $Y,$ and that the Lorenz force acting on it is balanced by the string tension. Eq.~(\ref{reflf}) arises from the
requirement of $N=1$ world-sheet supersymmetry.

It is easy to check that $R$ satisfies
$$
R^tGR=G,
$$
i.e. $R$ is an orthogonal transformation of $TX\vert_Y.$
This implies that on the boundary
the left-moving and right-moving $N=1$ supercurrents are equal:
$$
Q^++Q^-=\bQ^+ +\bQ^-.
$$
Thus such a boundary condition automatically preserves $N=1$ superconformal symmetry and therefore corresponds to a 
D-brane~\cite{Polchinsky}.

Boundary conditions for a topologically twisted 
sigma-model must in addition preserve $N=2$ 
superconformal symmetry~\cite{Witten}. This can be achieved in two
inequivalent ways: either we must have 
$$
Q^\pm=\bQ^\pm,\quad J=\bJ,
$$ 
or 
$$
Q^\pm=\bQ^\mp,\quad J=-\bJ,
$$ 
on the boundary. In the first case
we say that we have a B-type boundary condition, while in the second case we have an A-type boundary condition. One can show
that a B-type boundary condition corresponds to a B-brane, while an A-type boundary condition corresponds to an A-brane~\cite{Witten}.

It is easy to see that $R$ corresponds to
a B-type boundary condition if and only if $R^t\omega R=\omega.$ Since
$R$ is orthogonal, this is equivalent to saying that $R$ commutes with the complex structure $I=G^{-1}\omega.$ 
The latter condition obviously implies that $Y$ is a complex
submanifold in $X,$ and, less obviously, that 
$F$ is of type $(1,1).$ Thus a B-brane is a complex submanifold in $X$ with a holomorphic line bundle. This is the standard 
result~\cite{Witten,OOY}.

On the other hand, $R$ corresponds to
an A-type boundary condition if and only if
\begin{equation}\label{acond}
R^t\omega R=-\omega.
\end{equation}
To analyze this equation, let us choose a basis in $TX\vert_Y$ in which the
first $\dim_\RR X-\dim_\RR Y$ vectors span $NY$ and the remaining $\dim_\RR Y$
vectors span $TY.$
Let $\omega^{-1}$ have the following form in this basis
$$
\omega^{-1}=\begin{pmatrix} A & B \\ -B^t & C \end{pmatrix},
$$
where $A=-A^t,C=-C^t.$
Then the condition Eq.~(\ref{acond}) is equivalent to the following conditions on $A,B,C$:
\begin{eqnarray}
A &=& 0,\\
BF&=& 0,\\
gCg &=& FCF.
\end{eqnarray}

The first condition means that $Y$ is a coisotropic submanifold of $X.$
This implies that $\omega\vert_Y$ has a constant rank, 
and the dimension of the bundle ${\cL Y}
=\ker\left(\omega\vert_Y\right)$ is equal to the codimension of $Y.$

The second condition is equivalent to the statement that if we regard the 2-form $F$ as a bundle morphism $TY\ra TY^*,$ then its restriction to ${\cL Y}$ vanishes. In other words, if we denote by ${\cF Y}$ the quotient bundle $TY/{\cL Y},$ then $F$ descends to a section of $\Lambda^2 {\cF Y}.$ We will denote this section
$f.$ The form $\omega$ gives rise to another section of $\Lambda^2 
{\cF Y},$ which we will call $\sigma.$ Obviously, $\sigma$ is non-degenerate and makes ${\cF Y}$ into a symplectic bundle (i.e. a vector bundle with a smoothly varying symplectic structure on the fibers).

Now let us analyze the third condition. The metric $g$ provides a canonical
splitting $TY={\cL Y}\op {\cF Y},$ and it is easy to see that $C$ is simply
$0\op\sigma^{-1}.$ The K\"ahler property of the metric then implies
$$
gCg=0\op(-\sigma),
$$
and therefore the third condition is equivalent to
$$
f\sigma^{-1}f=-\sigma.
$$
In other words, if we denote the endomorphism
$\sigma^{-1}f:{\cF Y}\ra {\cF Y}$ by $J,$ then $J^2=-1.$
Thus ${\cF Y}$ has a natural complex structure.\footnote{Note that $\cF Y$
is both a complex bundle and a symplectic bundle, but it is not a unitary
bundle. The symplectic form $\sigma$ on the fibers has type $(0,2)+(2,0)$ in the complex structure $J.$ Thus $\sigma J=f$ is a skew-symmetric pairing,
rather than a K\"ahler metric.}

An obvious consequence of the first condition is that
$\dim_\RR Y - \frac{1}{2}\dim_\RR X$ is a non-negative integer. The other two
conditions imply that this integer is even. Indeed, the complex structure
$J$ leads to the Dolbeault decomposition of $\Lambda^2 {\cF Y},$ and it
is easy to
see that both $\sigma$ and $f$ are forms of type $(0,2)+(2,0).$ Since
both forms are non-degenerate, it follows that the complex dimension of
${\cF Y}$ must be even. This in turn implies that
$\dim_\RR Y - \frac{1}{2}\dim_\RR X$ is even.

For example, when $X$ is a 4-dimensional manifold ($T^4$ or a K3 surface),
an A-brane can be either 2-dimensional or
4-dimensional. When $X$ is 6-dimensional, an A-brane can be either 3-dimensional or 5-dimensional.
Note that a Calabi-Yau 3-fold which is a complete intersection in
a projective space has $H_5(X,\ZZ)=0,$
and therefore any 5-dimensional A-brane must be homologically trivial.
This seems to suggest that all A-branes are middle-dimensional in 
this case.

Let us consider two extreme cases. If
$\dim_\RR Y=\frac{1}{2}\dim_\RR X,$ then the first condition on $Y$ says that $Y$ is
Lagrangian. Since ${\cL Y}=TY$ in this case, the second condition says that
$F$ is zero, i.e. the line bundle $E$ is flat. The third condition is
vacuous in this case. Thus a middle-dimensional A-brane is a Lagrangian submanifold with a flat unitary line bundle. This is the standard result~\cite{Witten,OOY}

Another extreme case is $Y=X.$ In this case ${\cL Y}$ is the zero
vector bundle, and the first two
conditions are trivially satisfied. The bundle ${\cF Y}$ coincides with $TX,$
and thus the third condition says that $J=\omega^{-1}F$ is an almost complex
structure on $X$:
\begin{equation}\label{auxzero}
\left(\omega^{-1}F\right)^2=-id.
\end{equation}
We will see in the next section that $J$ is integrable,
and thus $X$ is a complex manifold. Note that $X$ has a complex structure $I$
to begin with, but the topological A-model is insensitive to it. Given
an A-brane wrapping the whole $X,$ one can construct a new
complex structure $J$ out of $\omega$ and $F.$ It is necessarily
different from $I,$ because $\omega$ has type $(1,1)$ with respect to $I$
and type $(2,0)+(0,2)$ with respect to $J.$

If $X$ is compact, the 2-form $F$ must have integer periods,
and it is clear that the equation $(\omega^{-1}F)^2=-id$ can be
satisfied
only for very special $\omega.$ For example, if $X$ is a 4-torus and
$\omega$ is generic, no line bundle on $X$ can be an A-brane. Presumably,
this implies that generically all A-branes are Lagrangian submanifolds in $X.$ But for some special $\omega$ there appear additional A-branes with $\dim_\RR Y=4.$

Let us show that this ``jumping'' phenomenon is mirror to the one described in Section~2. Recall that in Section~2 we considered
a complex torus $A$ of a very special kind ($n$\!\!-th power of an
elliptic curve with a complex multiplication, $n>1)$.
The Grothendieck group of $D^b(A)$ and its image in $H^*(A,\QQ)$ are unusually large. We also described a map $\beta$ from the rational
cohomology of $A$ to the rational cohomology of its
mirror $B,$ and showed that in general the image of $\beta$
does not lie in the middle-dimensional cohomology of the mirror
torus. For example, for $n=2$ the image of $\beta$ lies in the
even cohomology, and it can happen that $\beta$ maps
the Chern character of a coherent sheaf on $A$ to
an element which looks like the Chern character of a complex
vector bundle on $B.$ We interpreted this as saying that the mirror
of a coherent sheaf on $A$ can be a complex vector bundle
on $B.$ The Chern classes of such a vector bundle are not
arbitrary, but must satisfy certain constraints; for $n=2$ these
constraints are given by Eq.~(\ref{condchern}). When the rank of the
bundle is $1,$ we can compare these constraints with the algebraic
constraint on the curvature Eq.~(\ref{auxzero}).
The condition~(\ref{auxzero})
means that the rank of the matrix $F-i\omega$
is half the dimension of $X.$ If we set $\dim_\RR X=2n,$ then this
implies that $n$ is even, and that the $n/2+1$\!\!-st exterior
power of $F-i\omega$ vanishes. For $n=2$ the latter condition
is equivalent to
$$
F\wedge\omega=0, \qquad F\wedge F=\omega\wedge\omega.
$$
On the level of cohomology, these conditions are the same as
Eq.~(\ref{condchern}) in the special case $r=1.$
A similar argument can be made for $n>2.$

\section{The geometry of A-branes}

In this section we discuss the geometry of a general coisotropic A-brane. We will see that it has some beautiful connections with bihamiltonian
geometry and foliation theory.

A coisotropic submanifold $Y$ of a symplectic manifold $X$ has several equivalent
definitions. The usual definition is that at any point $p\in Y$
the skew-orthogonal complement of $TY_p$ is contained in $TY_p.$
Another popular definition is that $Y$ is
locally defined by first-class constraints. In other words, locally $Y$ can be represented as the zero-level of a finite set of smooth functions on $X$ all of whose Poisson brackets vanish on $Y.$

For our purposes, yet another definition will be useful. A submanifold
$Y$ is coisotropic if and only if
the restriction of $\omega$ to $Y$ has a constant rank, and
its kernel
${\cL Y}\subset TY$ is an integrable distribution.
This means that the commutator
of any two vector fields in ${\cL  Y}$ also belongs to ${\cL Y}.$

By the Frobenius theorem, this induces a foliation of $Y$ such that the vector
fields tangent to the leaves of the foliation are precisely the vector fields
in ${\cL Y}.$ The dimension of the leaves is equal to the codimension of $Y.$
We may call ${\cal L}Y$ the tangent bundle of the foliation.
The quotient bundle ${\cF} Y=TY/{{\cal L}Y}$ is called the
normal bundle of the foliation. (Elementary notions from foliation theory
that we will need can be found in Chapter 1 of Ref.~\cite{Tondeur}.)

If we interpret $Y$ as a first-class constraint surface in a phase space of
a mechanical system, then the meaning of the above foliation can be understood
as follows. First-class constraints lead to gauge symmetries. A leaf in $Y$
is precisely an orbit of a point under all gauge transformations. Formally,
the reduced phase space $Y_{red}$ describing gauge-invariant degrees of
freedom is the quotient of $Y$ by gauge transformations. In other words,
$Y_{red}$ is the space of leaves of the foliation. However, this space in
general does not have good properties, e.g. it need not be a manifold, or
even a Hausdorff topological space. Generally, it is unclear how to define dynamics on $Y_{red}.$

Instead, Dirac instructed us to work with gauge-invariant observables on $Y,$ i.e. with smooth functions on $Y$ which are locally constant along the leaves of the foliation. Such functions form a sheaf
$\cO_\cF(Y),$ which we can regard as the structure sheaf of the 
foliated manifold $Y.$ It plays the role of the
(generally non-existent) sheaf of smooth functions on the space $Y_{red}.$
Similarly, the sheaf of sections of ${\cF Y}$ locally constant along the leaves of the foliation replaces the tangent sheaf of $Y_{red}.$
We will denote this sheaf $\cT_\cF(Y).$

An A-brane is a coisotropic submanifold $Y$ with an additional structure:
a unitary line bundle $E$ on $Y$ whose curvature $F$
satisfies certain constraints. As explained in the previous section, this additional structure makes
${\cF Y}$ into a complex vector bundle with complex structure $J.$ It is easy to see that both $F$ and
$\omega$ are constant along the leaves, i.e.
$$
\cL_u F=\cL_u \omega =0,\quad \forall u\in \Gamma({\cL Y}).
$$
Thus $J=\sigma^{-1}f$ is also constant along the leaves. This means
that $J$ defines a transverse almost complex (TAC) structure on $Y.$ TAC structure is an analogue of almost complex structure for foliated manifolds. In the case when $Y_{red}$ is a manifold, giving a TAC structure on $Y$ is the same as giving an almost complex structure on $Y_{red}.$

The ``foliated'' analogue of a complex manifold is a manifold with a
transverse holomorphic structure (see e.g.~\cite{GM} for a definition and discussion). If $Y_{red}$ is a manifold, a transverse
holomorphic structure on $Y$ is simply a complex structure on $Y_{red}.$ In general, the definition goes as follows. A codimension $2q$ foliation on $Y$ is specified locally by a submersion $f:U\ra \RR^{2q}\simeq\CC^q,$ where $U$ is a coordinate
chart.\footnote{A submersion is a smooth map whose derivative is surjective.}
On the overlap of two charts $U$ and $V$ the two respective submersions $f$
and $g$ are related by a transition diffeomorphism
$\tau:f\left(U\cap V\right)\ra g\left(U\cap V\right).$
A transverse holomorphic structure on $Y$ is specified by a collection of
charts covering $Y$ such that all transition diffeomorphisms are bi-holomorphic.

The ``foliated'' analogue of the sheaf of holomorphic functions
is the sheaf of functions which are locally constant along the leaves and
holomorphic in the transverse directions. A remarkable feature of this sheaf
is that for a compact $Y$ all its cohomologies are
finite-dimensional~\cite{DK,GM}. Similarly, one can define
transversely holomorphic bundles on $Y,$ and again for compact $Y$ their sheaf
cohomologies are finite-dimensional~\cite{GM}. In general,
properties of compact transversely holomorphic manifolds are very similar to
those of compact complex manifolds.

It is easy to see that every transverse holomorphic structure
gives rise to a TAC structure. A TAC structure which arises in 
this way is
called integrable. The integrability condition for a TAC structure is the vanishing of the corresponding Nijenhuis torsion 
defined as follows. Let $u$ and $v$ be local sections of $\cT_\cF(Y).$ 
It is easy to see that the Lie bracket on $TY$ descends to a
Lie bracket on $\cT_\cF(Y),$ therefore the commutator $[u,v]$ is
well defined. The Nijenhuis torsion $T(J)$ 
is a section of $\cF Y\ot \Lambda^2\cF Y^*$ whose value on $u,v$
is defined to be
$$
T(J)=[Ju,Jv]-J[Ju,v]-J[u,Jv]+J^2[u,v].
$$
In the case of a trivial foliation, this reduces to the standard
definition of the Nijenhuis torsion of an almost complex structure.

Obviously, an integrable TAC structure has a vanishing Nijenhuis torsion, because in suitable coordinates $J$ is constant.
Conversely, by analogy with the classical case, one expects that any
TAC structure with a vanishing Nijenhuis torsion is integrable.
Indeed, as noted in Ref.~\cite{DK}, this is a special case of a 
theorem proved by Nirenberg~\cite{Nir}.
Thus there is a one-to-one correspondence between
transverse holomorphic structures on a foliated manifold $Y$ and
TAC structures on $Y$ with a vanishing Nijenhuis torsion.

A remarkable and non-obvious fact is that the TAC structure $J$ on an
A-brane $Y$ is automatically integrable. Let us give a proof of this fact
for the extreme case when $Y=X$ and the foliation is trivial
(i.e. each leaf is a point). It is easy to extend the proof to general
coisotropic A-branes.

First note that both $\omega$ and $F$ are symplectic structures on $X.$
Furthermore, since $\omega^{-1}F$ has eigenvalues $\pm i,$
$\omega_t=\omega+tF$
is non-degenerate for any real $t,$ and therefore is a symplectic structure as well. Hence its inverse is a Poisson structure for any real $t.$ Now note that by virtue of $(\omega^{-1}F)^2=-id$
the inverse has a very simple form
$$
\omega_t^{-1}=\left(1+t^2\right)^{-1}\left(\omega^{-1}+tF^{-1}\right).
$$
Thus any linear combination of $\omega^{-1}$ and $F^{-1}$ is a Poisson structure on $X.$ In the language of bihamiltonian
geometry~\cite{Dorfman,Magri},
$\omega^{-1}$ and $F^{-1}$ are compatible Poisson structures on $X.$ Now we can use the fundamental theorem of bihamiltonian geometry~\cite{Dorfman,Magri}
which says that if two Poisson structures $a$ and $b$ are compatible, and $a$ is non-degenerate, then the endomorphism $a^{-1}b:TX\ra TX$ 
has a vanishing Nijenhuis torsion. This theorem implies that the Nijenhuis torsion of $J$ vanishes, and therefore $J$ is integrable.

For a general coisotropic A-brane one can use the same argument,
but all objects are replaced by their foliated analogues: $TX$ is replaced by ${\cF Y},$ functions on $X$ are replaced by functions locally constant along
the leaves, Poisson structures are replaced by transverse Poisson
structures, etc. One can check that the fundamental theorem
of bihamiltonian geometry remains valid in the foliated case. In fact,
the version of this theorem proved in~\cite{Dorfman} (Theorem 3.12)
is valid in a very general setting, where the exterior differential complex
of a smooth manifold is replaced by an arbitrary complex over a Lie algebra.
The statement we need is a special case of this theorem.

We have shown that if there exists an A-brane with $Y=X,$ 
then $J=\omega^{-1}F$ is a complex structure on $X.$ Furthermore,
one can easily see that $F+i\omega$ is a closed
2-form on $X$ of type $(2,0)$ and maximal rank, i.e. a holomorphic
symplectic form. Thus in the complex structure $J$ the manifold $X$ is 
a compact holomorphic symplectic manifold. If in addition $X$ admits a K\"ahler metric compatible with $J,$ then $X$ is necessarily 
hyperk\"ahler~\cite{Bea}.
In general, $X$ need not be hyperk\"ahler for an A-brane with $Y=X$ to exist.

\section{A-branes and Homological Mirror Symmetry}

We have shown that an A-brane is a coisotropic submanifold in $X,$
and that it is naturally a foliated manifold with a transverse holomorphic structure. Now let us see how this fits in with the Homological Mirror Symmetry
Conjecture.

As explained in Section~\ref{intro}, 
the mirror of the derived category is
the category of A-branes. We have seen that in general the set of A-branes includes non-Lagrangian coisotropic branes, and therefore the Fukaya category must be enlarged with such A-branes for the
Homological Mirror Symmetry Conjecture to be true. Of course, in some
special cases there may be no non-Lagrangian A-branes, and the generalization we are proposing is vacuous. For example, there are no non-Lagrangian A-branes on an elliptic curve for dimensional reasons.
It also seems likely that there are no non-Lagrangian A-branes on
odd-dimensional Calabi-Yaus which are complete intersections in 
projective spaces, because any non-Lagrangian A-brane would be
homologically trivial. 
Nevertheless, we believe that a uniform formulation
of the Homological Mirror Symmetry Conjecture for all weak Calabi-Yau manifolds would be very illuminating. Let us see how one far one can go in this direction.

One immediately sees the following major difficulty.
A Lagrangian A-brane can carry a flat vector bundle of rank $r$ higher
than one. From a physical viewpoint, such an A-brane should be
thought
of as $r$ coincident A-branes of rank one. The same reasoning suggests that there
exist coisotropic A-branes with higher rank bundles. However, it is not clear
to us what the constraints on the connection are in this case, and whether
a transverse holomorphic structure arises again. Thus we do not really
understand all the objects in the enlarged Fukaya category.

We will ignore this difficulty and try instead to say something about
morphisms between the objects we already know. Unfortunately, understanding
morphisms between different A-branes is not much easier than understanding
A-branes with higher rank bundles: the former question is just an
``infinitesimal'' form of the latter. Therefore we will focus on the
{\it endomorphisms} of coisotropic A-branes.

To guess the right definition, let us look at the two extremes: Lagrangian
A-branes and A-branes wrapping the whole $X$ (i.e. $Y=X)$. The space of endomorphisms of a Lagrangian A-brane $Y$ is its Floer homology
$HF_*(Y,\CC).$ This is hard to compute, but in many cases it coincides
with the de Rham cohomology $H^*(Y,\CC).$ From a physical
viewpoint, the de Rham cohomology is a classical approximation to the
Floer homology; the two coincide when there are no world-sheet instanton
contributions to the path integral computing the Floer
differential~\cite{Witten}.

Now suppose we have an A-brane $Y=X.$ This means that there exists
a unitary line bundle on $X$ with a connection 1-form $A$ whose curvature
$F=dA$ satisfies
\begin{equation}\label{aux}
(\omega^{-1}F)^2=-id.
\end{equation}
As explained in the previous section, this implies
that $J=\omega^{-1}F$ is a complex structure on $X.$ On general grounds,
endomorphisms of an A-brane must have the structure of a graded vector space (in physical terms, the grading is given by the ghost charge).
A natural guess is the Dolbeault cohomology $H^{0,*}(X)$ with respect to $J.$

As a simple check, note that degree one elements in the space of
endomorphisms must parametrize infinitesimal deformations of the A-brane.
In the present case, a deformation is a real 1-form $a$ such that the curvature
of the connection 1-form $A+a$ satisfies Eq.~(\ref{aux}) up to terms
quadratic in $a.$ This is equivalent to the condition
$$
(da) J + J^t(da)=0,
$$
i.e. $da$ must be a form of type $(1,1).$ If we denote by $a''$ the
$(0,1)$ part of $a,$ then the latter condition is equivalent to
$\bpartial a''=0.$
Thus $a''$ represents a class in $H^{0,1}(X).$ 
Since $a$ is real, the $(1,0)$ part of $a$ is determined by $a''$
(is complex conjugate to it). Thus there is a natural map from
the space of deformations of an A-brane to $H^{0,1}(X).$ 

We want to show that this map becomes one-to-one, if we quotient the 
space of deformations by deformations which are isomorphisms in the category of A-branes. Obviously, the usual infinitesimal
gauge transformations $a=df,$ where $f$ is a real function on $X,$ induce isomorphisms. However, this is not all. 
In the case of Lagrangian A-branes it is known that a flow along a Hamiltonian vector field on $X$ induces an isomorphism in the Fukaya category, and it is
natural to assume that the same is true for more general coisotropic A-branes. 
If $h$ is a smooth real function on $X,$ and $V_h=\omega^{-1}dh$ is the corresponding Hamiltonian vector field, then the induced deformation of
the connection 1-form $A$ on $X$ is
$$
a=\cL_{V_h} A = i_{V_h} F+ d\left(i_{V_h} A\right),
$$
where $\cL_V$ is the Lie derivative along $V.$
Thus the most general deformation $a$ which is an isomorphism in the
category of A-branes has the form
$$
a=i_{V_h} F + df,
$$
where $h$ and $f$ are arbitrary smooth real functions on $X.$ 
Taking into account the relation $J=\omega^{-1}F,$ this can be rewritten as
$$
a=-2 J^t dh +df=\partial (f+2i h) +\bpartial (f-2ih).
$$
Let us denote by $\Ext^1$ the space of deformations of the A-brane
modulo isomorphisms.
It is easy to check that the map from the
space of deformations to $H^{0,1}(X)$ descends to a well-defined map 
from $\Ext^1$ to $H^{0,1}(X),$ and that the latter map
is an isomorphism of real vector spaces, as claimed.

With these two examples in mind, it is not hard to guess the right graded
vector space for a general coisotropic A-brane. If $Y$ is a foliated
manifold with a transverse holomorphic structure, recall
that we denoted by
$\cO_\cF(Y)$ the sheaf of complex functions on $Y$ which are locally constant along the leaves of the foliation and holomorphic in the transverse directions. We propose that the space of endomorphisms of 
a coisotropic A-brane $Y$ is the cohomology of the sheaf $\cO_\cF(Y).$

It is trivial to see that our proposal is consistent with the two extreme
cases considered above. For a Lagrangian A-brane, $\cO_\cF(Y)$ 
is simply the
sheaf of locally constant complex functions on $Y,$ and its cohomology
coincides with the de Rham cohomology of $Y.$ 
For $Y=X$ $\cO_\cF(Y)$ is the sheaf of holomorphic functions on $X$ (with respect to the complex structure $J)$, 
and we again get agreement.

It would be very interesting to understand how morphisms between
different coisotropic A-branes should be defined. At first sight, no
suitable complex whose cohomology one could compute presents itself. Perhaps this is simply a lack of imagination on our part.

In general, it appears that a geometric definition of the category
of A-branes is very cumbersome. Finding such a definition is akin to trying to
define the category of holomorphic vector bundles on a complex manifold
using the zeros of their holomorphic sections.
A more promising approach is to look
for an algebraic definition of A-branes, for example as modules over some
non-commutative algebra associated to a symplectic manifold $X.$ It seems
likely that this non-commutative algebra is related to the
deformation quantization of $X.$ Similar ideas have been discussed
in~\cite{NT,Soi}.

\section*{Acknowledgments}
We are grateful to Dan Freed for useful suggestions and to Ezra Getzler for pointing out the relevance of bihamiltonian geometry. 
Some preliminary results have
been presented by A.K. at the Duality Workshop, ITP, Santa Barbara,
June 18 -- July 13, 2001.
A.K. would like to thank Ron Donagi, Dan Freed, Ezra Getzler,
Tony Pantev, and other participants for stimulating discussions,
and the organizers for making this workshop possible. A.K. was
supported in part by DOE grants DE-FG02-90ER40542 and DE-FG03-92-ER40701.
D.O. was supported in part by RFFI grant 99-01-01144 and
a grant for support of leading scientific groups N 00-15-96085.
The research described in this publication was made possible in part
by Award No RM1-2089 of the U.S. Civilian Research and Development
Foundation for the Independent States of the Former Soviet Union (CRDF).

\end{document}